RESEARCH ARTICLE

# Characterizing College Science Assessments: The Three-Dimensional Learning Assessment Protocol

James T. Laverty[1¤a]*, Sonia M. Underwood[2¤b], Rebecca L. Matz[1], Lynmarie A. Posey[2], Justin H. Carmel[2], Marcos D. Caballero[1,3], Cori L. Fata-Hartley[4], Diane Ebert-May[5], Sarah E. Jardeleza[1¤c], Melanie M. Cooper[1,2]

1 CREATE for STEM Institute, Michigan State University, East Lansing, Michigan, United States of America, 2 Department of Chemistry, Michigan State University, East Lansing, Michigan, United States of America, 3 Department of Physics and Astronomy, Michigan State University, East Lansing, Michigan, United States of America, 4 College of Natural Science, Michigan State University, East Lansing, Michigan, United States of America, 5 Department of Plant Biology, Michigan State University, East Lansing, Michigan, United States of America

¤a Current address: Department of Physics, Kansas State University, Manhattan, Kansas, United States of America
¤b Current address: Department of Chemistry & Biochemistry and STEM Transformation Institute, Florida International University, Miami, Florida, United States of America
¤c Current address: University Analysis, Reporting, and Assessment, Salisbury University, Salisbury, Maryland, United States of America
* laverty@ksu.edu

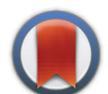

**OPEN ACCESS**

**Data Availability Statement:** All relevant data are within the paper and its Supporting Information files.

**Funding:** The Leona M. and Harry B. Helmsley Charitable Trust (http://helmsleytrust.org/) awarded the Association of American Universities (http://www.aau.edu/) Grant #2012PG-EDU005. Michigan State University received a sub-award of that grant (MMC). The funders had no role in study design, data collection and analysis, decision to publish, or preparation of the manuscript.

## Abstract

Many calls to improve science education in college and university settings have focused on improving instructor pedagogy. Meanwhile, science education at the K-12 level is undergoing significant changes as a result of the emphasis on scientific and engineering practices, crosscutting concepts, and disciplinary core ideas. This framework of "three-dimensional learning" is based on the literature about how people learn science and how we can help students put their knowledge to use. Recently, similar changes are underway in higher education by incorporating three-dimensional learning into college science courses. As these transformations move forward, it will become important to assess three-dimensional learning both to align assessments with the learning environment, and to assess the extent of the transformations. In this paper we introduce the Three-Dimensional Learning Assessment Protocol (3D-LAP), which is designed to characterize and support the development of assessment tasks in biology, chemistry, and physics that align with transformation efforts. We describe the development process used by our interdisciplinary team, discuss the validity and reliability of the protocol, and provide evidence that the protocol can distinguish between assessments that have the potential to elicit evidence of three-dimensional learning and those that do not.

# Introduction

From "A Nation at Risk" to "the Gathering Storm" to "Engage to Excel", there is a clear need to improve STEM education for both the preparation of future STEM professionals and for the development of a scientifically literate citizenry [1–4]. In response to these reports, several national level initiatives in K-12 and higher education were devised and implemented. For example, three sets of science education standards were developed for K-12 science education, the most recent being the Next Generation Science Standards (NGSS) [5–8]. While there is not a coherent approach to the design and implementation of curriculum reform in higher education, there are a number of influential reports. The recent "Discipline-Based Education Research" report summarized the current state of science and engineering education in higher education, described the existing research on education in science and engineering fields, and specified areas of research that should be expanded [9]. A following report, "Reaching Students," discussed research-based classroom pedagogy strategies for practitioners [10]. Other national- level initiatives include the "Vision and Change" report for undergraduate biological sciences and the reinvention of Advanced Placement (AP) courses in biology, chemistry, and physics [11–15].

To date, much of the research on transformation of higher education STEM courses has focused on incorporating research-based pedagogies in college classrooms, laboratories, and discussion sections. It is now clear that "active learning" can improve overall course grades and retention in STEM courses, particularly for underprepared and underrepresented students [16], but what is less clear is what students are learning in these courses. There is ample evidence that students (even graduate students) emerge from STEM courses (even those using "active learning" techniques) with profound misunderstandings of important ideas in many disciplines, and there is little evidence that students are able to transfer their knowledge to new situations [9]. Our work considers not only what students should know, but also how they know it and how they can use that knowledge. We are considering ways in which both the curriculum and assessment of student learning should be transformed in addition to incorporating research-based pedagogical strategies.

We (and others) have previously argued that one way forward is to use the best available synthesis of evidence about student learning and the resultant vision for the future of science education [17,18]. Advocates suggest that we use the National Research Council's (NRC) report "A Framework for K-12 Science Education: Practices, Crosscutting Concepts, and Core Ideas" (herein referred to as the Framework) to help us restructure our science curricula, instruction, and the ways we assess student learning in higher education [19]. The Framework supports the idea of building scaffolded curricula around progressions of disciplinary core ideas, using scientific practices, and emphasizing crosscutting concepts (herein referred to as three-dimensional learning, and defined in more detail below). At Michigan State University (MSU), we are engaged in such a transformation effort. Here we report the development of a protocol to determine the extent to which assessments from gateway courses in biology, chemistry, and physics provide opportunities for students to engage with the three dimensions defined in the Framework. This protocol can be used to evaluate assessment tasks, characterize entire assessments, and guide faculty in developing three-dimensional assessment tasks.

## Three-Dimensional Learning

The Framework synthesized much of the literature about how students learn science into a vision for how science and engineering education should be implemented. The three dimensions outlined in the Framework are:

- *What students should be able to do with their knowledge*. The Framework describes eight **"scientific and engineering practices"** that can be thought of as the disaggregated components of inquiry. These include ideas such as *Developing and Using Models*, *Constructing Explanations*, *Planning and Carrying out Investigations*, and *Engaging in Argument from Evidence*. It is precisely these practices that can engage students in the processes of science and engineering, by using their knowledge to model systems, explain phenomena, and design solutions to problems.

- *Ideas common across scientific disciplines*. The Framework identified seven **"crosscutting concepts"** that span science disciplines. The concepts include ideas such as *Cause and Effect*, *Systems and System Models*, and *Structure and Function*. These concepts can be thought of as "ways of thinking" about phenomena that have the potential to allow students to make connections across disciplines.

- *Concepts essential to the study of a discipline*. The third dimension in the Framework suggests that students should develop their knowledge around **"disciplinary core ideas"** rather than try to assemble their understanding of important ideas from a large number of disparate facts and fragments of knowledge. We know that disciplinary experts have a great deal of knowledge at their disposal, and it is organized and contextualized around a few important concepts [20]. By focusing learning around the core ideas of the discipline, students are given more support to make connections among the concepts and build a more robust framework of knowledge. Evidence suggests that disciplinary core ideas should be developed over time with carefully scaffolded progressions of learning activities and formative assessments [20].

The authors of the Framework emphasized that the three dimensions should not be taught separately, but should instead be integrated in all aspects of learning, which includes instruction, curriculum, and assessment.

While the Framework was written for a K-12 audience, the findings almost certainly apply to college classrooms, as there is little reason to believe that students completing high school learn in a different way than beginning college students (who, in many cases, were completing high school a few months earlier).Recent changes to AP courses have shifted to the idea of blending scientific practices and core ideas [13–15]. Additionally, both the DBER Report and Vision and Change have highlighted elevating scientific practices to the same level as content [9,11]. The primary reason for the restriction of the Framework to K-12 was that the committee's charge was limited to the K-12 education system. The approach outlined by the Framework contrasts sharply with the topics-driven curriculum aligned with chapters in textbooks currently found in many college science classrooms.

Drawing from the Framework, we have argued that the idea of three-dimensional learning has the power to address a number of shortcomings that exist in higher education [17]. It allows departments to address the "mile wide, inch deep" problem by first determining the centrally important ideas and then focusing the curriculum on those goals [21]. Using the Framework as a guide for curriculum development or revision has the potential to change the conversation from asking questions about what can be removed from the curriculum, to what must we teach for students to achieve the goals we have set for them, a fundamental idea in backward design [22].

By identifying and focusing on the centrally important concepts within (core ideas) and across (crosscutting concepts) the disciplines, we can help students build a robust understanding of science. This organization of understanding around core ideas mirrors how experts think about their disciplines [20]. Additionally by engaging students with these concepts in the same ways that scientists do (scientific practices), students learn to apply their knowledge to new situations, a key goal in higher education [9]. This approach allows students the opportunity to develop an understanding of the nature of evidence, models, and explanations in science, capabilities that can serve them well in a world where scientific competence is increasingly necessary [23,24]. Engagement with all three dimensions also allows students to gain insight not only into the concepts that scientists deem important, but also into why scientists decided these concepts are important in the first place, which has implications for affecting students' epistemology toward science [25].

The research presented here focuses on how course assessments in biology, chemistry, and physics can be used to elicit evidence of student engagement with the three dimensions of the Framework, which requires identifying characteristics of assessment tasks that align with certain criteria.

## Focusing on Assessment

The Framework was designed to provide a vision for how science and engineering education could build on accumulated research and evidence from the learning sciences. It also provided the basis for the NGSS, operationalized as a set of performance expectations of increasing sophistication over time. Each performance expectation provides a three-dimensional description of what students should be able to do with their knowledge [8]. The transformation efforts at MSU follow this approach by working to blend the scientific practices, crosscutting concepts, and core ideas into the gateway science courses. To incorporate more opportunities for three-dimensional learning in our courses, we must communicate to our students the importance of these ideas, which means we must change the ways that student performance is assessed [26,27]. The assessments must match the expectations we have for students, meaning that assessment systems focused on the three dimensions must be developed.

At the introductory level in higher education, assessments tend to focus on factual recall, procedural knowledge, and algorithmic problem solving [28]. While there have been a number of approaches to changing the types of assessments used, none address three-dimensional learning. For example, beginning in the late 1980's with the development of diagnostic tests, followed by the more famous concept inventories (e.g. the *Force Concept Inventory* in physics, the *Chemistry Concepts Inventory* in chemistry, and the *Genetics Concepts Assessment* in biology), attempts to develop selected response tests based on research on student thinking resulted in multiple choice questions about concepts along with distractors that encompass typical incorrect student ideas or misconceptions [29–32]. These tests and inventories are productive for investigating when students learn more scientifically normative ideas or concepts [16,33]. However, these types of assessments were never designed to determine if students can use these ideas in the same ways that scientists do, nor do they assess whether students can use their knowledge. In short, while experts certainly can do well on these tests, doing well on such a test does not provide evidence that students can use those ideas to predict or explain phenomena (i.e. to do science). Therefore, while diagnostic tests and concept inventories are important and have proven useful in the past, their appropriateness for assessing three-dimensional learning is likely limited and should be evaluated (see Discussion).

Another approach to the development of assessments in order to provide information about student understanding is to invoke Bloom's taxonomy [34,35]. This approach involves designing questions to align with a set of hierarchical levels using action verbs such as "select" for low levels and "construct" for upper levels. Indeed, Bloom's taxonomy is widely used, especially in post-secondary biology education research and faculty professional development as well as in chemistry and physics [36–41]. However, the meaning of Bloom's descriptors can be unclear (particularly from the perspective of students completing the assessments), fine distinctions between the levels are difficult to justify, and faculty value more than just the level of a question [42,43]. Further, recent reports of research on science teaching and learning do not include Bloom's taxonomy as a useful construct for developing learning objectives or assessing student learning [9,11,19,44,45]. Yet Bloom's taxonomy is useful to clarify the distinction between learning facts and deeper conceptual understanding. However, the scientific practices outlined in the Framework are based on observation and analysis of actual scientific practice, suggesting that they can be used to more systematically characterize what we want students to know and be able to do with that knowledge.

More recently, the NRC report "Knowing What Students Know" proposed the idea of assessment as an evidentiary argument [46]. An assessment should elicit evidence about the construct that is being assessed, and from that evidence, an argument about student learning can be constructed. One approach to assessment as an evidentiary argument is Evidence-Centered Design [47]. In this approach, one defines a construct to be assessed, decides what is convincing evidence that students understand this construct, and designs assessment tasks that can elicit this evidence. The student responses can then be analyzed using a range of methods, such as using a rubric or applying item response theory. The result is an iterative process that guides the development of assessment tasks that address the outcomes of interest

Our work treats assessment as an evidence-based argument; the constructs of interest are the Framework's three dimensions and our objective is to be able to recognize assessment tasks that can elicit evidence of three-dimensional learning. As we discuss later, our intent is to assess the *potential* of assessment tasks to elicit such evidence; what we cannot do is evaluate the quality of the evidence that might be produced from that task, because we cannot know what students will do when faced with such tasks. Evaluation of the evidence is left to the developers of the assessment tasks.

Although the NRC report "Developing Assessments for the Next Generation Science Standards" [8,45] summarizes existing literature and provides guidance for assessment developers (including recommending Evidence-Centered Design), few examples of three-dimensional assessment tasks are currently available. Notably, the authors of that report state that selected response tasks should not be the sole method of collecting information about student learning, as there is ample evidence that selected response tests overestimate the ability of students to use their knowledge [48]. The NRC report further concludes:

> "CONCLUSION 2–1 [. . .] assessment tasks will generally need to contain multiple components (e.g., a set of interrelated questions). It may be useful to focus on individual practices, core ideas, or crosscutting concepts in the various components of an assessment task, but, together, the components need to support inferences about students' three-dimensional science learning. . ." [45]

Ultimately, the report's authors recommend that a combination of constructed response tasks, selected response tasks, and projects should be used to assess three-dimensional learning. Much of this advice is directed towards assessment developers, or offered as a guide for developing formative assessments in small classrooms. However in college level science courses, faculty, typically disciplinary experts without formal experience in assessment design, write most assessments. Particularly in large enrollment gateway courses, selected response assessment tasks are most often used to assess fragmented bits of knowledge and basic skills (e.g. dimensional analysis). If transformation efforts are to succeed, the assessments used in courses must reflect the transformation goals. It is important to characterize how existing and future assessment tasks have the potential to elicit evidence that the student engaged with a scientific practice, crosscutting concept, or core idea (for simplicity, the phrase "have the potential to elicit evidence that the student engaged with" will herein be shortened to "elicit"). We also will need to provide guidance for faculty about ways to develop three-dimensional assessments. To help meet these needs, we developed the Three-Dimensional Learning Assessment Protocol (3D-LAP), which describes criteria for each scientific practice, crosscutting concept, and core idea that define what is required for an assessment task to elicit that particular construct. Potential uses for the 3D-LAP include 1) the identification of individual assessment tasks as three-dimensional, 2) the characterization of course assessments, and 3) faculty development.

Our goal was to develop a protocol that allows users to identify assessments that can elicit evidence of three-dimensional learning. In this paper, we describe how we developed the protocol and provide evidence that the protocol can distinguish between assessments that can elicit the ideas in the Framework and those that do not.

## Methods

### The Three-Dimensional Learning Assessment Protocol (3D-LAP)

The development process began with a review of the Framework, NGSS, and supporting literature [8,19,49–52]. In the initial phase of this work, the development team (consisting of disciplinary experts, many of whom identify as discipline-based education researchers, in biology, chemistry, and physics) focused its discussions on the scientific practices and crosscutting concepts as outlined in the Framework (Table 1). The engineering-specific practices defined in the Framework were excluded from consideration because of our interest in developing a protocol that was applicable to assessments in science courses (and because we didn't want to overreach our expertise, which does not include engineering). Representation of experts from biology, chemistry, and physics on the development team was essential for arriving at an understanding of what it means for an assessment task to elicit a particular scientific practice that is applicable across disciplines. In addition, the very nature of crosscutting concepts demands input from across these three disciplines to reach a consensus on their key features that can be used to identify their presence in assessments.

In parallel, members of the development team and colleagues from their respective disciplines began identifying and negotiating the core ideas of each discipline. Inclusion of faculty beyond the development team in this part of the process provided an opportunity to consider more diverse perspectives in the identification and elaboration of core ideas representative of the views of disciplinary faculty at our institution. Furthermore, development of the 3D-LAP was part of a larger initiative to transform the introductory science courses in each discipline [53]. Faculty from each discipline identified a set of core ideas and criteria that assesses those ideas. We anticipate that users of the 3D-LAP may choose to identify their own set of core ideas and criteria, but provide our lists as examples of how the ideas can be operationalized.

**Table 1. The scientific and engineering practices and crosscutting concepts as listed in the Framework [19].**

| Scientific and Engineering Practices | Crosscutting Concepts |
|---|---|
| 1. Asking Questions (for science) and Defining Problems (for engineering)<br>2. Developing and Using Models<br>3. Planning and Carrying Out Investigations<br>4. Analyzing and Interpreting Data<br>5. Using Mathematics and Computational Thinking<br>6. Constructing Explanations (for science) and Designing Solutions (for engineering)<br>7. Engaging in Argument from Evidence<br>8. Obtaining, Evaluating, and Communicating Information | 1. Patterns<br>2. Cause and Effect: Mechanism and Explanation<br>3. Scale, Proportion, and Quantity<br>4. Systems and System Models<br>5. Energy and Matter: Flows, Cycles, and Conservation<br>6. Structure and Function<br>7. Stability and Change |

doi:10.1371/journal.pone.0162333.t001

Initially, we used the descriptions of scientific practices and crosscutting concepts found in the Framework to develop assessment criteria for determining whether an assessment task elicits a particular scientific practice or crosscutting concept. These criteria were applied to a range of assessment tasks, including textbook questions, tasks from reformed curricula, and tasks we determined exemplified particular scientific practices, crosscutting concepts, or core ideas. By using these assessment tasks, we were able to describe a range of tasks that faculty use for examinations and refine the criteria. Discrepancies between our negotiated understandings of the scientific practices, crosscutting concepts, and core ideas, and the evaluation of assessment tasks using the criteria promoted revisions of the criteria. This process was repeated until the development team was satisfied that the criteria were consistent with our negotiated understandings of the three dimensions. The result of this process takes a different form for each of the three dimensions: each scientific practice includes a list of 2–4 criteria, all of which must be met in order for a task to elicit that scientific practice; each crosscutting concept has a brief description of what is necessary for a task to elicit that crosscutting concept; and each core idea comes with a list describing ideas that must be involved in the task to qualify as eliciting a core idea (only one item from the list must be aligned to count). The protocol itself can be found in S1 Protocol.

Throughout the development process, we encountered examples of assessment tasks that satisfied some but not all of the criteria for a particular scientific practice, or did not fit the description of a particular crosscutting concept or core idea. This was especially common for scientific practices that include reasoning as a central element, such as *Developing and Using Models* and *Constructing Explanations*. Despite meeting most criteria for the practice, the reasoning component was often absent in assessment tasks. To distinguish between assessment tasks that fully met the criteria for a particular scientific practice and those that satisfied most of the criteria, we experimented with "explicit" and "implicit" subcategories. Likewise, we characterized the presence of a particular core idea (or crosscutting concept) as "explicit" or "implicit" based on whether the assessment task had the potential to get students to use the core idea (or crosscutting concept) or was simply related to the core idea (or crosscutting concept). However, it became clear that this approach was ineffective because the ultimate goal is to help faculty construct assessment tasks aligned with three-dimensional learning. Hence coding assessment tasks that could, at best, invoke "implicit" evidence are not meaningful in this context. Therefore, the possible classification of assessment tasks as "implicit" or "explicit" was abandoned as an element of the 3D-LAP.

Below, we discuss the development of the criteria for each of the three dimensions separately and the implications for developing and characterizing assessments.

## Scientific Practices

Development of the criteria for identifying the scientific practices in assessments was guided by the question "What specific components must be present in a task in order for it to potentially elicit evidence of student engagement with a given practice?" The development team considered what it means to engage in each of the eight scientific practices (Table 1) defined in the Framework and negotiated an understanding of what each of the practices entails. The engineering-specific practices of defining problems and designing solutions were not considered in development of the 3D-LAP as our focus was assessments in biology, chemistry, and physics.

During the development of the 3D-LAP, it became apparent that separate criteria were needed to determine if an assessment task elicited a scientific practice depending on whether students were asked to construct or select a response. While assessment tasks that require a constructed response provide the strongest evidence of three-dimensional learning, large enrollment college courses and a concomitant lack of resources for scoring often necessitate the use of selected response (e.g. multiple choice) tasks. Because the scientific practices typically require that students "construct", "develop", or "engage", we developed parallel criteria for the scientific practices for both constructed response and selected response assessment tasks. The development team both recognizes and supports the idea that selected response tasks are not sufficient to provide strong evidence of student engagement with scientific practices as noted in the NRC report, "Developing Assessments for the Next Generation Science Standards",

> "Traditional approaches [. . .] which heavily rely on selected-response items, will not likely be adequate for assessing the full breadth and depth of the NGSS performance expectations, particularly in assessing students' proficiency with the application of the scientific and engineering practices in the context of disciplinary core ideas." [45]

Separate criteria were developed for constructed and selected response tasks because the ways in which students can engage with the scientific practices differ significantly. Table 2 compares the criteria for *Developing and Using Models* to illustrate the differences between constructed- and selected-response tasks that elicit scientific practices. Selected response tasks that meet these criteria do not provide strong evidence of three-dimensional learning, however, we included these criteria because selected responses are widely used in assessments.

Adaptation of the scientific practices presented in the Framework became necessary as we defined what students would be asked to do to provide evidence of engaging with the practices on assessments and then actually applied the criteria developed to assessment tasks. For example, we found significant overlap in the criteria generated for the scientific practices of *Constructing Explanations* and *Engaging in Argument from Evidence*. It was sufficiently difficult to produce criteria that clearly distinguished between the two practices over the range of assessment tasks analyzed across the three disciplines that we ultimately merged the two sets of criteria into a single practice in the 3D-LAP: *Constructing Explanations & Engaging in Argument from Evidence*. The distinction between explanation and argumentation is unclear as researchers in the field continue to discuss whether or not explanations and arguments need to be distinguished [49,54,55].

The 3D-LAP does not distinguish between summative and formative assessments but is applicable to both types. Hence, we adapted several other scientific practices. For instance, *Obtaining, Evaluating, and Communicating Information* was reduced to simply *Evaluating Information* because in the summative assessment tasks collected from introductory biology, chemistry, and physics lecture courses, we found no examples where students were asked to

**Table 2. Constructed and selected response criteria for the scientific practice of *Developing and Using Models* from the 3D-LAP.** Differences are highlighted in bold.

| Constructed Response Criteria | Selected Response Criteria |
|---|---|
| 1. Question gives an event, observation, or phenomenon for the student to explain or make a prediction about. | 1. Question gives an event, observation, or phenomenon for the student to explain or make a prediction about. |
| 2. Question gives a representation or asks student to **construct** a representation. | 2. Question gives a representation or asks student to **select** a representation. |
| 3. Question asks student to **explain or make a prediction** about the event, observation, or phenomenon. | 3. Question asks student to **select an explanation for or prediction** about the event, observation, or phenomenon. |
| 4. Question asks student to **provide the reasoning** that links the representation to their explanation or prediction. | 4. Question asks student to **select the reasoning** that links the representation to their explanation or prediction. |

doi:10.1371/journal.pone.0162333.t002

*Obtain* information from an external source. We also were unable to produce exemplar tasks for *Obtaining Information* suitable for summative assessment. We eliminated *Communicating* from the practice because, arguably, every response could be construed as communicating information, although not necessarily in keeping with the vision of the Framework. *Planning and Carrying Out Investigations* was reduced to *Planning Investigations* because it was unlikely that students in a lecture course would be asked to carry out an investigation as part of a summative assessment task. Finally, we removed *Asking Questions* from the scientific practices for selected response tasks because we found no tasks nor could we construct tasks that addressed this practice in a meaningful way. The most closely related task asked, "Which of the following is a scientific question?" Such a task, however, provides no evidence that students can generate productive scientific questions when presented with data, observations, or phenomena. Our adaptations of the scientific practices outlined in the Framework reflect differences in purpose between the Framework and the 3D-LAP. The scientific practices and their descriptions in the Framework are intended to guide curriculum development, instruction, and both formative and summative assessment, whereas the 3D-LAP was developed as a tool to identify the potential of summative and formative assessment tasks to engage students in three-dimensional learning.

The final list of scientific practices in the 3D-LAP and the associated criteria can be found in S1 Protocol.

## Crosscutting Concepts

In the initial negotiations of the crosscutting concepts, we followed an approach similar to that for the scientific practices. We defined criteria for each of the crosscutting concepts found in the Framework (Table 1) based on descriptions found therein. These criteria were used to characterize existing and new assessment tasks across the three disciplines and were then iteratively revised based on feedback as the development team applied them to assessment tasks. As with the practices, it was important to reach agreement among the disciplines about the meaning of each crosscutting concept and how that meaning would be operationalized. Over time, it became clear that using a list of criteria, all of which must be satisfied to identify the presence of a crosscutting concept (similar to the scientific practices), was limiting. There were multiple instances of assessment tasks that members of the development team thought aligned with a particular crosscutting concept yet failed to meet all of the criteria. This prompted the development team to rethink the appropriateness of using a list of criteria to define the knowledge encompassed by crosscutting concepts that span disciplines.

In contrast with the scientific practices, which provide ways for students to use their knowledge (actions), the crosscutting concepts are elements of that knowledge with explanatory value across the disciplines (ideas). We ultimately developed brief descriptions of what is needed to elicit each crosscutting concept to help coders identify components that should be present. No differentiation was made in the descriptions for constructed and selected response tasks. An example of these descriptions is

To code an assessment task with Structure and Function, the question asks the student to predict or explain a function or property based on a structure, or to describe what structure could lead to a given function or property.

We made one modification to the crosscutting concepts found in the Framework (Table 1). In the Framework, the ideas of scale, proportion, and quantity are combined in a single crosscutting concept: *Scale*, *Proportion*, and *Quantity*. The development team determined that the concept of scale is sufficiently different from the ideas associated with proportion and quantity to merit independent consideration. Hence, we separated scale from proportion and quantity to create two independent crosscutting concepts in the 3D-LAP: *Scale* and *Proportion and Quantity*. The complete set of criteria for the crosscutting concepts developed for the 3D-LAP is found in S1 Protocol.

## Core Ideas

A different approach from the one used for the scientific practices and crosscutting concepts was required for the development of the core ideas criteria in the 3D-LAP. In the cases of scientific practices and crosscutting concepts, we were able to use those found in the Framework with minor adaptations. In contrast, the disciplinary core ideas for the Physical Sciences and the Life Sciences in the Framework, even for grade bands 9–12, do not align directly with the content and organization by discipline of many university science courses. Consequently, the development team determined that new lists of the centrally important ideas for each discipline were needed. In the 3D-LAP, we refer to these as "core ideas" to distinguish them from the "disciplinary core ideas" contained in the Framework. The requirements that our core ideas should satisfy were informed by the criteria outlined for the disciplinary core ideas on page 31 of the Framework. In order for a concept to qualify as a core idea in the 3D-LAP, it must have 1) disciplinary significance, 2) the power to explain a wide range of phenomena, and 3) potential for generating new ideas.

Development of the 3D-LAP was part of a larger effort to transform the gateway science courses at our institution, and as such, it was important to engage a broad sample of disciplinary faculty beyond the development team, particularly those involved in teaching these courses. Shifting the focus in assessment from what students know to how students use their knowledge to explain and predict phenomena begins with discussions among faculty about what knowledge they value. Disciplinary faculty were invited to participate in defining the core ideas for their discipline so that the resulting product would be valued and potentially used by faculty beyond the development team as we try to expand the incorporation of three-dimensional learning into assessment and instruction [56]. In all three disciplines, groups of faculty at MSU worked together to reach agreement on the core ideas within each discipline, guided in part by national efforts [11,13–15]. Members of the 3D-LAP development team participated in this process as peers. The ways in which faculty outside the 3D-LAP development team were involved in identifying and defining cores ideas played out differently in each discipline: One discipline had broad participation from its faculty in discussions about the core ideas; another created committees responsible for the two gateway courses that worked to develop the core ideas; the third discipline had a small working group develop the core ideas and then solicited feedback from disciplinary faculty. Examples of one core idea developed by each discipline are provided in Fig 1. In order for a task to elicit a core idea, the task must align with either the main definition or one of the bullets under it.

While these core ideas are useful to our transformation effort and to the work of the development team, we readily accept and encourage the possibility that other groups may develop and use a set of core ideas that is relevant to their project or discipline. As with the crosscutting concepts, the criteria with which we characterized the core ideas are descriptive because they have a negotiated meaning within their respective disciplines. Research demonstrates that disciplinary experts are able to readily identify the underlying principles that govern an assessment task; principles from which most of our core ideas stem [57]. This is likely a result of disciplinary experts having negotiated the meaning of these concepts as part of the process of joining their community of practicing scientists [58]. Our complete set of core ideas and their criteria are found in S1 Protocol.

**Biology**

5. **Structure and Function:** The functions and properties of ecosystems, organisms, tissues, cells, and biological molecules are determined by their structures.
   ○ At the molecular level, biology is based on dynamic, three-dimensional chemical and physical interactions.

**Chemistry**

1. **Electrostatic and Bonding Interactions**: Attractive and repulsive electrostatic forces govern noncovalent and bonding (covalent and ionic) interactions between atoms and molecules. The strength of these forces depends on the magnitude of the charges involved and the distances between them.
   ○ Attractive noncovalent interactions (intermolecular forces) between atoms and molecules arise from interactions between transient, induced, and permanent dipoles.
   ○ Atoms also interact through electrostatic forces to form chemical bonds, which have greater stability (lower energy) than the separated atoms. Nonpolar covalent bonding and ionic bonding represent the limits in a continuum of bonding interactions with polar covalent bonds falling in-between. Electrons in the highest energy orbitals with the largest spatial extent (valence electrons) are used to form bonds.
   ○ Covalent and ionic bonding interactions are typically stronger than noncovalent (intermolecular) interactions.

**Physics**

1. **Interactions Can Cause Changes in Motion**: Changes in an object's motion are the result of interactions between it and one or more other objects. Multiple interactions between an object and its surroundings can result in a predictable change in motion.
   ○ All macroscopic forces are the result of gravitational or electromagnetic interactions between particles.
   ○ Non-zero net forces cause changes in linear momentum.
   ○ Non-zero net torques cause changes in angular momentum.
   ○ Changes in linear and angular momentum can be used to predict the motion of objects.

**Fig 1. Examples of the criteria found in the 3D-LAP for core ideas, one each from biology, chemistry, and physics.**

doi:10.1371/journal.pone.0162333.g001

## Using the 3D-LAP

### Characterizing Assessments

The first goal for the 3D-LAP is to characterize existing assessment tasks in biology, chemistry, and physics courses. An assessment task coded as eliciting a scientific practice, must meet all

**A)**

**Scientific Practice: Developing and Using Models**

*2. Construct a representation*                                    *4. Connect representation to explanation*

*Create a diagram* that shows the molecular structure of the lipid bilayer in a typical cell membrane. *Use the diagram* to explain why **oxygen ($O_2$) can easily pass through the membrane but sodium ions ($Na^+$) cannot.**

*3. Explanation*                          **1. Phenomenon**

**B)**

**Crosscutting Concept: Structure and Function**
**Core Idea: Structure and Function**

**Structure**

Create a diagram that shows the **molecular structure of the lipid bilayer** in a typical cell membrane. Use the diagram to explain why *oxygen ($O_2$) can easily pass through the membrane but sodium ions ($Na^+$) cannot.*

*Function*

**Fig 2. Assessment task from an introductory biology course that elicits evidence of three-dimensional learning.** Panel A shows the parts of the task that meet each of the criteria for the *Developing and Using Models* scientific practice (see Table 2). Panel B shows the parts of the task that meet the criteria for both the crosscutting concept (*Structure and Function*, given in the text above) and core idea (*Structure and Function*, see Fig 1). Further analysis of this task is provided in the S1 Supporting Information.

doi:10.1371/journal.pone.0162333.g002

the criteria for a particular practice. Assessment tasks coded as eliciting a crosscutting concept, must align with the brief description given for the crosscutting concept. Assessment tasks coded as eliciting a core idea, must align with either the primary description or one of its sub-bullets. Fig 2 shows a constructed response assessment from introductory biology that was characterized as eliciting each of the three dimensions (i.e. it is a "three-dimensional task"). Numerous examples (13 from each discipline) of coded assessments from biology, chemistry, and physics are provided in the S1 Supporting Information.

In each of the disciplines, there is at least one core idea that is similar to a crosscutting concept (such as the ones in Fig 2). In these cases, there was some concern that certain tasks would "elicit two dimensions for the price of one". The development team determined that disentangling the criteria for the crosscutting concept from the criteria for the core idea (and vice versa) was unnecessary because a task that elicits one can certainly elicit the other. In each case of overlap between a crosscutting concept and core idea, the criteria are different between the two, so it is possible for tasks to be coded with one and not the other, though they are often coded together.

Examining each assessment task on an exam (or similar assessment) using the 3D-LAP allows us the opportunity to characterize entire assessments (our second goal). Fig 3 shows a representation of two chemistry exams that were coded using the 3D-LAP. These exams were specifically chosen because exam A was made up of assessment tasks like those found in most textbooks, while exam B was taken from a course that employed a research-based curriculum. These exams were chosen specifically because they were expected to emphasize the utility of the 3D-LAP: to identify assessments that have the potential to elicit evidence of three-dimensional learning. When comparing the Chemistry A and B exams, the differences detected by

**Fig 3. Comparison of two exams characterized using the 3D-LAP.** The first row of each diagram shows the question number. In the last three rows, blue, green, and red shaded cells, indicate there is evidence for a scientific practice, crosscutting concept, or core idea, respectively. Questions 21–23 on the Chemistry B exam are constructed response. All other questions shown are selected response.

doi:10.1371/journal.pone.0162333.g003

the 3D-LAP are apparent. The Chemistry A exam has few tasks that elicit one or more dimensions whereas the Chemistry B exam has more examples of tasks coded with one to three dimensions, with several coded with all three dimensions.

Characterizing assessments in this way can be used to compare courses over time (e.g., to gather evidence that a course transformation was successful), to compare exams within a course (e.g., to characterize what is being assessed in a course), and to compare exams between courses within a discipline (e.g., to characterize where students are assessed on each scientific practice within a degree program).

As part of our analysis, we tracked the percentage of total points associated with each task in addition to the codes we obtained using the 3D-LAP. We argue that the number of points assigned to certain tasks on an exam by instructors informs students about the relative importance of those tasks. We use this relative importance to demonstrate what percentage of points on each exam elicits the three dimensions. Fig 4 shows the difference between two exams we coded in each discipline. Like those in Fig 3, these exams were specifically chosen because the A exams were made up of canonical assessment tasks, while the B exams were taken from research-based curricula to emphasize the utility of the 3D-LAP. The exams shown in Fig 3 are the same exams shown in the Chemistry portion of Fig 4, now scaled by the number of points assigned to each task. Fig 4 shows that in each of the A exams, most of the points on the exam are given to tasks with little potential to elicit more than one dimension. In contrast, each of the B exams is dominated by tasks that might elicit evidence of two or three dimensions, with at least 40% of the points being awarded for three-dimensional tasks.

Fig 5 shows another representation of the data, specifying the percentage of points on each exam that are associated with tasks that might elicit each of the dimensions. We found that in the A exams, all three disciplines award few points for tasks that elicit any dimension, with the exception of the Physics A exam, where core ideas appeared in many assessment tasks. The most likely explanation for the difference is that the physics core ideas are more fine-grained that the biology and chemistry core ideas. As was discussed above, the core ideas were intentionally developed to be meaningful and align with the discipline at our institution. Faculty in physics argued that the governing principles that appear in introductory physics (e.g., Newton's second law, Conservation of Energy) were prevalent in many assessment tasks including those that asked students to solve canonical numerical problems. However, they conceded that few problems of that type gave students opportunities to engage with scientific practices or crosscutting concepts. Across all three disciplines, we see that the B exams assign substantially more points to tasks that elicit scientific practices, crosscutting concepts, and core ideas.

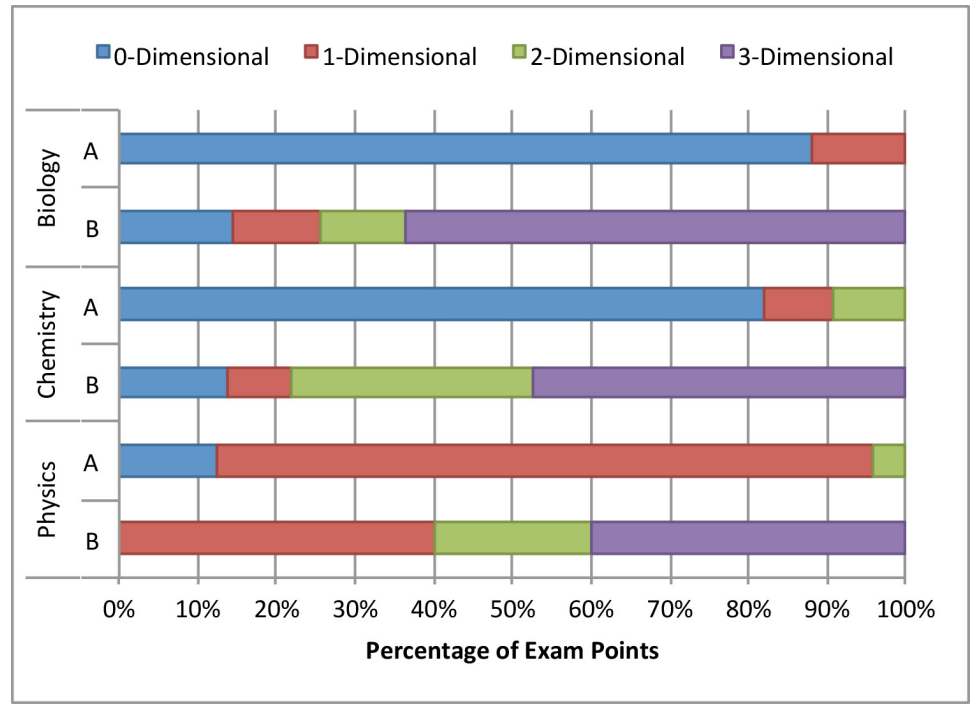

**Fig 4. Comparison of the percentage of exam points assigned to tasks that were coded with each of the three dimensions for two exams per discipline.** The Chemistry A and B exams are the same ones shown in Fig 3.

## Validity and Reliability

Here we establish the validity and reliability of the protocol for the purposes of characterizing assessments as part of the transformation efforts at MSU. It is important to note that the protocol itself should not be considered valid and reliable for all possible datasets, because validity and reliability are characteristics of analysis tools and data, together [59,60]. Therefore, we coded exams from our dataset to investigate the validity and reliability for this project. Groups conducting other studies should establish validity and reliability for their datasets. We provide a model for doing so below.

Many types of validity and reliability testing are available, but not all are appropriate for the 3D-LAP. We describe the evidence that supports the face and content validity of the criteria for our data. In the case of the 3D-LAP, our team of disciplinary experts and discipline-based education researchers in biology, chemistry, and physics developed the protocol together, which allowed us to operationalize the criteria across all three disciplines. Extensive negotiation took place as part of an iterative cycle among the development team to develop the criteria for each of the scientific practices and crosscutting concepts. When creating the criteria for the practices and crosscutting concepts, both theory (e.g. the Framework) and practice (i.e. existing assessment tasks) were considered at each phase of negotiation. A similar process was established with the disciplinary faculty teams (in addition to the development team), which worked to ensure that all aspects of the core ideas were included. For these reasons, we claim that the protocol with our dataset has both face validity and content validity.

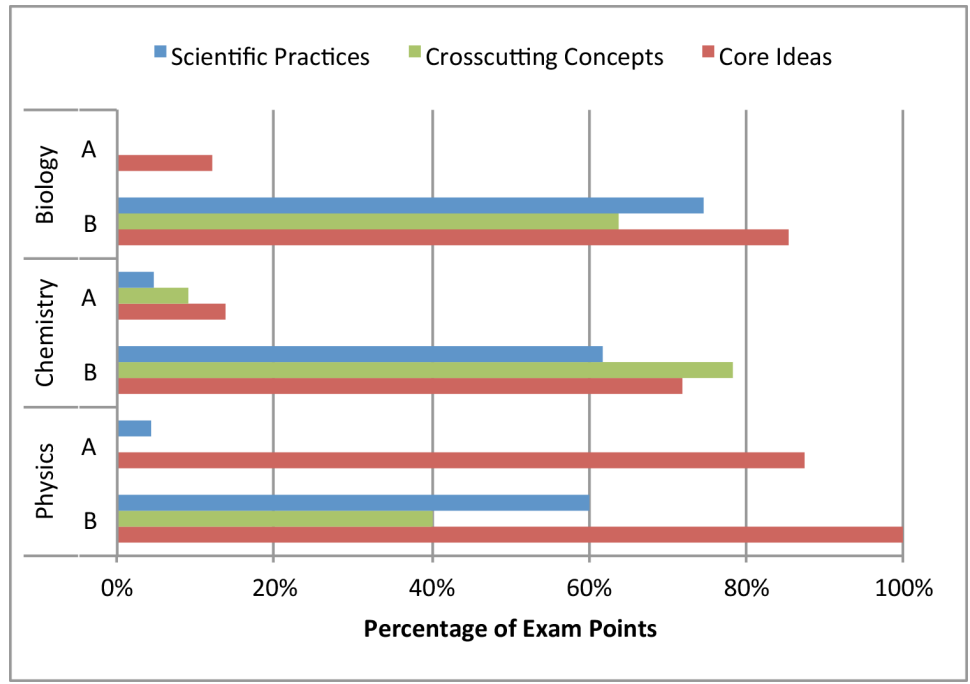

**Fig 5. Comparison of two exams from each discipline, displaying the percentage of exam points assigned to tasks coded with scientific practices, crosscutting concepts, and core ideas.** This representation shows that although there are few zero dimensional tasks in physics, the vast majority of the tasks address core ideas, and the other two dimensions are almost never elicited.

doi:10.1371/journal.pone.0162333.g005

To assess the reliability of the protocol with our data, we coded two exams per discipline from our dataset with representative assessment tasks. These exams were specifically chosen because we expected one exam to include very few tasks that elicit the three dimensions and the other exam to include many more. Researchers from the development team coded each task in their discipline and percent agreement was used to evaluate inter-rater reliability. For this comparison, we considered two coders to be in agreement as long as they both indicated that the task could elicit evidence of a scientific practice, crosscutting concept, or core idea, or agreed that they did not ([Table 3](#) shows a summary of cases). We chose this definition of agreement because tasks can meet the criteria for multiple scientific practices, crosscutting concepts, or core ideas. For example, asking students to construct the chemical structure for a set of compounds followed by ranking and explaining the boiling point trend for the compounds meets the criteria for both *Constructing Explanations* & *Engaging in Argument from Evidence* and *Developing and Using Models*. In this instance, coders may agree that the assessment task could be coded as either of the two practices; however, unless the agreement is taken at a higher level ("is there a practice?") the coders disagree if one coded both practices and another selected just

**Table 3. Possible coding of tasks by two coders and the resulting code for inter-rater reliability.**

|  | Coder 1 | Coder 2 | Result |
|---|---|---|---|
| Scenario 1 | Developing and Using Models | Developing and Using Models | Agree |
| Scenario 2 | No Practice | No Practice | Agree |
| Scenario 3 | Developing and Using Models; Evaluating Information | Developing and Using Models | Agree |
| Scenario 4 | Developing and Using Models | Planning Investigations | Agree |
| Scenario 5 | Developing and Using Models | No Practice | Disagree |

doi:10.1371/journal.pone.0162333.t003

**Table 4. The percent agreement values for each of the three dimensions, which we use to determine our inter-rater reliability using the 3D-LAP to evaluate assessment tasks in our dataset.**

| Discipline | # of coders | # of tasks | Scientific Practices (%) | Crosscutting Concepts (%) | Core Ideas (%) |
|---|---|---|---|---|---|
| Biology | 3 | 63 | 95 | 93 | 95 |
| Chemistry | 4 | 48 | 86 | 78 | 77 |
| Physics | 2 | 31 | 93 | 87 | 87 |

doi:10.1371/journal.pone.0162333.t004

one (Scenario 3 in Table 3). Therefore, disagreements were only defined when one coder said that the task elicited a scientific practice, crosscutting concept, or core idea, and another coder determined that it did not (Scenario 5 in Table 3).

After coding the exams, we measured inter-rater reliability by looking at the percent agreement between the coders within the discipline, setting our lower limit at 75%. For biology and chemistry (the disciplines with >2 coders), the percent agreement was calculated for each pair of coders and those values were averaged. Table 4 shows the number of coders for each discipline, the total number of tasks coded in each discipline, and the values for the percent agreement for each dimension in each discipline.

## Discussion

The 3D-LAP can be used to characterize the potential of assessment tasks to elicit evidence that students engage with scientific practices, crosscutting concepts, and core ideas in the context of college level introductory biology, chemistry, and physics courses. The protocol can be used reliably by researchers and differentiates exams that have the potential to elicit evidence of three-dimensional learning from those that do not. This new protocol addresses the NRC report, "Developing Assessments for the Next Generation Science Standards", which states, "new kinds of science assessments are needed to support the new vision and understanding of students' science learning" [45]. The 3D-LAP helps guide practitioners and researchers alike in how to identify or author these new kinds of science assessments.

We demonstrated that the 3D-LAP can be used to characterize tasks as having the potential to elicit evidence that the student engaged with a scientific practice, crosscutting concept, and/or core idea. By repeating this process for every task on an assessment (such as a homework set or exam), we can use the 3D-LAP to make comparisons between assessments. Such comparisons can be useful to evaluate the success of a transformation effort by comparing assessments before and after the transformation, to identify what students are assessed on in a particular course, and to investigate where students are assessed on various aspects of three-dimensional learning as part of a degree program.

We are using the 3D-LAP to assess transformation efforts at MSU. By characterizing assessments from both before and after the transformation, we will investigate whether or not the transformation resulted in assessments that can elicit scientific practices, crosscutting concepts, and core ideas. Additionally, the 3D-LAP will be used to explore the impact of three-dimensional learning on students' recruitment and persistence in STEM fields.

We are also investigating the utility of existing assessments (e.g. concept inventories) for the purposes of evaluating students' three-dimensional learning. By identifying or modifying existing assessments or developing new assessments, we can better assess students' abilities to engage in science using the core ideas of the discipline. We can begin to develop new types of assessment tasks to measure student learning in higher education, which in turn can be used to guide the development of instruction and curriculum. Ideally, students would achieve the goals of three-dimensional learning at both local and national levels.

We do not claim that all assessment tasks need to elicit all three dimensions, but if we engage in course transformations where three-dimensional learning is central, then we must also ensure that the course assessments align accordingly. To that end, we argue that assessments similar to the A exams in Figs 3, 4 and 5 do not assess students' abilities to engage in science the way that scientists do. However, we recognize that instructors may want to directly assess particular skills, such as drawing Lewis structures, constructing a Punnett square, or generating a free-body diagram (see S2 Supporting Information for more examples of such skills). Tasks assessing these kinds of skills alone are often coded as being zero- or one-dimensional using the 3D-LAP. Given the need to assess skills, coupled with the fact that higher-dimensional tasks take more time to answer, it is necessary to think about balance when designing an assessment. We make no claims about the optimal mix of tasks that makes an assessment "good," but do claim that an assessment should not be comprised almost entirely of zero- or one-dimensional tasks (such as the A exams), but rather an array of multi-dimensional tasks and skills (more similar to the B exams).

The 3D-LAP can also be used as a professional development tool to aid faculty in writing assessment tasks capable of eliciting evidence that a student engaged with a scientific practice, crosscutting concept, and/or core idea. Faculty interested in bringing the ideas of three-dimensional learning to their classroom must have ways to assess this kind of learning. The 3D-LAP can serve as a guide for faculty to develop assessment materials on their own, but they will likely need to be trained to use the protocol in ways that align with the designers' intent. We have run workshops at major conferences using the 3D-LAP in just this way. In two recent workshops (one in physics, the other in chemistry), the participants were introduced to the 3D-LAP, then asked to code existing assessment tasks, and finally used it to develop their own tasks. Anecdotally, the response to the instrument is positive and the use by novice users aligns with our intentions. Although workshops include a self-selected group of participants, we intend to determine the extent to which other faculty need guidance in applying and using the 3D-LAP, both for identifying tasks that align with three-dimensional learning and for developing new assessments.

The development of the 3D-LAP is based upon the existing research on student science learning and assessment. More work needs to be done to explore how students engage with and respond to three-dimensional assessments. Such work may result in modifications to the 3D-LAP or may reveal discipline-specific patterns that may need to be addressed in the future.

## Limitations

The 3D-LAP provides information about the potential of assessment tasks to elicit evidence of three-dimensional learning but does not consider how students actually respond to assessment tasks. Furthermore, we have not taken into account particular features of assessment tasks that influence how students respond, such as the structure of the prompt or question bias [46]. Therefore, tasks that we identify as potentially eliciting evidence of students engaging with the three dimensions may, but do not necessarily elicit three-dimensional thinking. Also, we do not infer what faculty intended for their assessment tasks, nor do we know the goals of courses where these assessment tasks are given. Finally, the 3D-LAP was developed for use in evaluating assessments from introductory college-level biology, chemistry, and physics lecture courses. Although we expect some components of the 3D-LAP to apply to other levels and types of courses, the 3D-LAP has not been tested in those courses.

## Conclusion

We developed the 3D-LAP, a protocol to identify assessment tasks that have the potential to elicit evidence of three-dimensional learning. To demonstrate its utility, we characterized six exams from our dataset; two each from biology, chemistry, and physics. These characterizations allowed us to identify assessments that have the potential to elicit evidence of what students can do with their knowledge of science.

The 3D-LAP was developed primarily to evaluate the success of a transformation effort that is focused on bringing three-dimensional learning into introductory courses in biology, chemistry, and physics at MSU. By characterizing how assessments change over time, we intend to evaluate the success of those efforts in future publications. We provided evidence that the protocol is both valid and reliable when applied to our dataset.

While the 3D-LAP was designed with assessments in introductory biology, chemistry, and physics courses in mind, it could be used in (or adapted to) other higher education courses or disciplines, or even to K-12 science courses. Future work with the protocol will include characterizing the extent to which our transformation efforts are successful, investigating how students engage with assessments that were developed guided by the 3D-LAP, and using the protocol as a professional development tool.

Finally, we include the full 3D-LAP and examples of characterizing tasks from biology, chemistry, and physics in S1 Protocol and S1 Supporting Information and encourage others to use it to characterize or develop their own assessments.

## Supporting Information

**S1 Dataset. Data.**
(XLSX)

**S1 Protocol. 3D-LAP.**
(DOCX)

**S1 Supporting Information. Exemplars.**
(DOCX)

**S2 Supporting Information. Skills.**
(DOCX)

## Acknowledgments

This project was supported by the Association of American Universities' Undergraduate STEM Education Initiative (which is funded by the Helmsley Charitable Trust) and by the Office of the Provost of Michigan State University. We thank the faculty who provided assessment tasks for our review, engaged in discussions of core ideas, and have taken steps to incorporate three-dimensional learning into their courses.

## Author Contributions

**Conceptualization:** JTL SMU RLM LAP JHC MDC CLF DE SEJ MMC.

**Data curation:** JTL SMU RLM LAP JHC MDC CLF DE SEJ MMC.

**Formal analysis:** JTL SMU RLM LAP JHC MDC CLF DE SEJ MMC.

**Methodology:** JTL SMU RLM LAP JHC MDC CLF DE SEJ MMC.

**Project administration:** JTL SMU RLM MMC.

**Resources:** JTL SMU RLM LAP JHC MDC CLF DE SEJ MMC.

**Supervision:** JTL SMU RLM MDC MMC.

**Validation:** JTL SMU RLM LAP JHC MDC CLF DE SEJ MMC.

**Visualization:** JTL SMU RLM LAP JHC MDC CLF DE SEJ MMC.

**Writing – original draft:** JTL SMU RLM LAP JHC MDC CLF DE SEJ MMC.

**Writing – review & editing:** JTL SMU RLM LAP JHC MDC CLF DE SEJ MMC.